# CCDs for the instrumentation of the Telescopio Nazionale Galileo

R. COSENTINO[1,3], G. BONANNO[1], P. BRUNO[1], S. SCUDERI[1], F. BORTOLETTO[2,3], M. DALESSANDRO[2], C. BONOLI[2], D. FANTINEL[2]

[1]Osservatorio Astrofisico di Catania, Catania, Italy
[2]Osservatorio Astronomico di Padova, Padova, Italy
[3]Telescopio Nazionale Galileo, Canary Island, Spain

ABSTRACT. Most of the scientific instrumentation as well as the tracking systems and the Shack-Hartmann wavefront analysers at the Italian National Telescope Galileo use CCDs as detectors. The characterization of detectors is of fundamental importance for their correct utilization in scientific instrumentation. We report on the measurement of the electro-optical characteristics of CCDs that will be used in the scientific instrumentation at the Italian National Telescope. In particular we will show and compare the quantum efficiency, the charge transfer efficiency, the dark current, the read out noise, the uniformity and the linearity of two sets of CCDs manufactured by EEV and LORAL. Finally, we will show the preliminary tests done at the telescope with the optical imager that has a mosaic of two EEV chips.

## 1. Introduction

It is no secret that CCDs are the dominant detectors in the instrumentation for optical astronomy. The Italian National Telescope is no exception to this rule. Three out of four of its instruments as well as the cameras for tracking and for the adaptive optics use CCDs as detectors. The measurement of the electro-optical characteristics of these detectors is the necessary preliminary step to allow the proper selection of the detector and then its optimal use at the telescope. We have characterized two sets of CCDs manufactured by EEV and LORAL selected for the instrumentation of the Italian National Telescope. In section 2 we will describe shortly the Italian National Telescope and its scientific instrumentation using CCDs. In section 3 we will outline the main characteristics of the CCD controller. Section 4 is devoted to the description of the facilities used for the calibration of the CCDs. Section 5 deals with the results of the measurements of the CCDs characteristics and finally in section 6 the results of few tests performed at the telescope are shown.

## 2. The Italian National telescope Galileo

The Telescopio Nazionale Galileo (TNG) is the national facility of the Italian astronomical community, and is located at Roque de los Muchachos (alt. 2400 m.) in La Palma (Canary Islands, Spain). TNG is an altazimuth telescope with a Ritchey-Chretien optical configuration and a flat tertiary mirror feeding two opposite Nasmyth foci. The diameter of the primary mirror is 3.58 m with a focal length of 38.5 m (F/li). The scale is 5.36 arcsec/mm and the vignetting-free field of view is 25 arcminutes in diameter. The telescope will host an optical and an infrared camera at Nasmyth focus A and two spectrographs at the Nasmyth focus B.

### 2.1. TNG Scientific Instrumentation using CCDs

As said before CCDs are widely used at TNG for guiding, tracking, adaptive optics and of course for the science instruments. The CCDs whose characteristics we have measured are however just the ones used in the scientific instrumentation. These instruments are: the optical imager, the low-resolution spectrograph and the high-resolution spectrograph.

#### 2.1.1. Optical Imager Galileo

The Optical Imager Galileo (OIG) is the CCD camera for the direct imaging at optical wavelengths (3200 – 11000 Å) for Galileo. It is placed at the Nasmyth adapter interface A and is usually illuminated by light coming directly from the TNG tertiary mirror to the f/11 focus, with no other optical elements in front of the CCD apart those on the filter wheels (filters, polarizers, and atmospheric dispersion correctors) and the dewar window. The OIG is designed to host a variety of CCD chips or mosaics covering a field of view up to 10 arcmin.

#### 2.1.2. Low Resolution Spectrograph

The Low Resolution Spectrograph (LRS) is a focal reducer instrument installed at the spectrographic Nasmyth focus (Nasmyth adapter interface B) of the TNG. The LRS camera is opened to F/3.2. The scale is 0.276 arcsec/pixel with a field of view of 9.4 square arcmin. Up to now the observing modes available are: direct imaging, long slit spectroscopy and multi-object spectroscopy. The camera is equipped with a 2Kx2K CCD LORAL.

#### 2.1.3. Spettrografo ad Alta Risoluzione Galileo

The Spettrografo ad Alta Risoluzione per Galileo (SARG) is the white pupil cross dispersed echelle spectrograph under construction for the TNG (Gratton et al 1998). SARG is a high efficiency spectrograph designed for a spectral range between 3700 and 9000 Å and with a resolution ranging from R=19,000 up to R=144,000. SARG uses an R4 echelle grating in Quasi-Littrow mode. The beam size is 100 mm giving an RS product of RS=46,000 at the center of the order. Single object and long slit (up to 30 arcsec) observing modes are possible.

### 3. The CCD controller

The CCD controller which runs all the CCDs at the TNG is described in detail in Bonanno et al. 1995 and Bortoletto et al. 1996. Here we will give a brief description of its main characteristics.

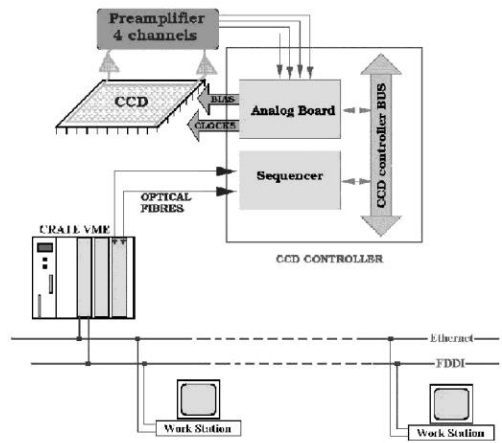

Fig. 1. CCDs readout system.

3.1. The control system of the CCD cameras.

The architecture of the CCDs readout system is shown in Fig 1. The first block is the CCD controller, which is located close to the cryostat. Into the CCD controller there is a Bus (the CCD controller bus), in which are plugged-in a sequencer board and at least one analog board. The sequencer board generates the clock sequences. Each analog board produces 8 programmable bias voltages, 16 clock drivers with independently programmable upper and lower levels and is also able to read and process four channels independently. The controller is connected to a VME crate through optical fibers. The VME contains a shared memory for image storing, a transputer to send commands to the controller and an Ethernet connection to the TNG workstation system.

 The acquisition system allow different readout modes: frame transfer (tracking cameras), full frame (scientific cameras), binning on chip (1X2, 2X2 ... 3X3 ) and read of boxes.

3.2. The Waveform Editor.

To allow different readout modes, the CCD controller generates table-based sequences. These tables are generated in an easy way by using a windows program. This program, named waveform editor, allows to generate the tables through a graphical interface and mouse actions. Fig. 2 shows an example of a waveform editor session.

4. Calibration facilities.

The main calibration of the CCDs used for the TNG instrumentation is performed at the Catania Astrophysical Observatory. The whole range of the CCDs electro-optical parameters can be measured using the facilities described in the following paragraph. At the telescope itself we have realized a small laboratory that allows to tune-up the CCDs with the electronics that makes it work at the telescope and to check the temporal behavior of their performances.

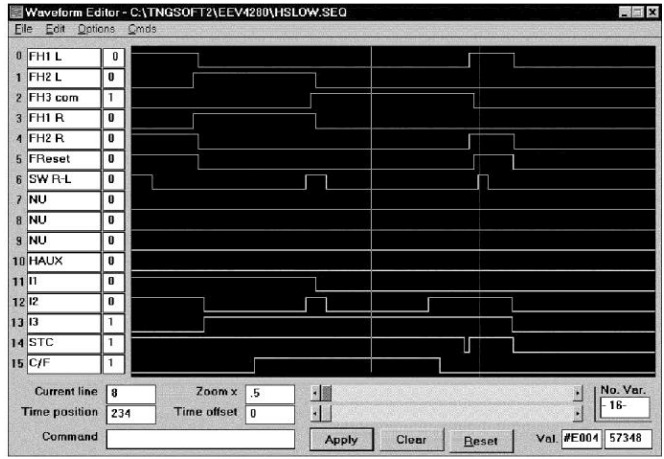

Fig. 2. Waveform Editor.

4.1. The calibration facility of the Catania Astrophysical Observatory

The Catania Astrophysical Observatory calibration facility allows a full electro-optical characterization of CCD detectors. The main component of the facility is the instrumental apparatus to measure the quantum efficiency in the wavelength range 1300 – 11000 Å. The radiation emitted by two light sources (deuterium and xenon lamp) goes through a series of diaphragms and filters and then is focused onto the entrance slit of a monochromator. The dispersed radiation beam then is divided by a beam splitter and focused on a reference detector and on the CCD. A detailed description of this apparatus can be found in Bonanno et al 1996. For uniformity and linearity measurements a 20 inches integrating sphere is optically connected to the QE measurements system through a quartz singlet. The useful wavelength interval range in this case is 2000 – 11000 Å. Finally, the system gain and the CTE measurements are performed using a $Fe^{55}$ x-ray source.

4.2. The calibration facility at the telescope

The calibration facility at the TNG detectors laboratory consists in a xenon lamp, two filter wheels with a series of interference filters and a series of neutral filters, a 20 inches integrating sphere and a reference photodiode.

5. CCDs Calibration.

The CCDs used in TNG instrumentation are manufactured by EEV and LORAL. All the chips are thinned back illuminated with UV enhanced response. The LORAL chips are thinned at the Steward Observatory and their UV response is enhanced using a technique developed there called the chemisorption (Lesser and Venkatraman, 1998). The EEV chips are ion implanted. Table 1 summarize the characteristics of the four CCDs (2 EEV and 2 LORAL) whose electro-optical performances have been measured. The two EEV and one LORAL CCD are being used at the OIG while the other LORAL will be used at the LRS. For each CCD, we have measured the system gain, the readout noise, the linearity, the uniformity, the quantum efficiency, the dark current and the charge transfer efficiency.

5.1. System gain

The calculation of the system gain, K (e-/DN), is done using the x-ray stimulation technique. We used a $Fe^{55}$ x-ray source that emits mainly 5.9 keV photons. The x-rays produce an ideal point source of charge of known energy. On average 3.65 eV of energy is needed to produce a single e-h pair yielding an average of 1620 e- produced in silicon per impinging x-ray photon. If the x-ray event occupies a single pixel is referred to as a single pixel event. Events that occupy more than a single pixel are referred to as split and/or partial events.

Tab. 1 - Manufacturer characteristics of TNG CCD

| Manufacturer | Loral | EEV |
|---|---|---|
| Chip | Thinned Back illuminated | Thinned Back illuminated |
| Pixel size | 15 μ | 13.5 μ |
| Area (pixel) | 2048 X 2048 | 2048 X 4096 |
| MPP | NO | Yes |
| Working temp. | -110 C | -130 C |
| UV treatment | Chemisorption | Ion implantation |
| AR Coating | YES | YES |
| Grade | 2 (both) | 2-3 |

5.2. Read-Out noise

The read-out noise is measured as the RMS of the signal in the overscan region. The read-out noise turned out to be 12 e- and 10 e- for the LORAL and the EEV chips respectively.

5.3. Linearity

The linearity is measured using a standard method of illuminating the CCD at different signal levels with a uniform source of radiation. The deviation from linearity measured for the two sets of CCD are 0.15 % for LORAL chips and 0.5 % for EEV chips.

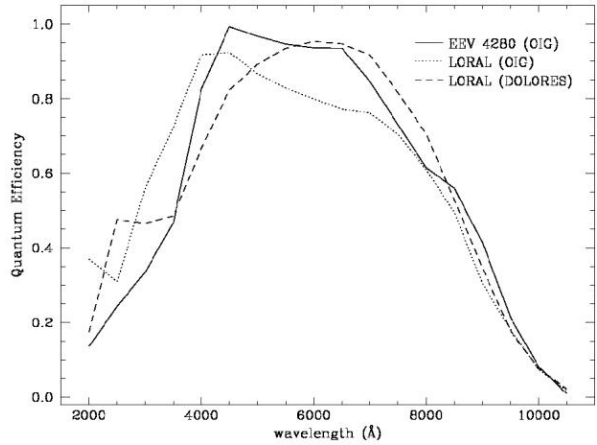

Fig. 3. QE measured in the wavelength 2000 10500 ~~

5.4. Uniformity

The homogeneity of the CCD is measured illuminating it with a uniform source of radiation at different wavelengths. The deviation from homogeneity is given then by the following expression:

$$DeviationfromHomogeneity = \frac{RMS(wholearea)}{mean(wholearea)} \%$$

Table 2 summarizes the deviation from uniformity for a LORAL and an EEV chip

Tab. 2 - Deviation from uniformity

| λ (Å) | Loral | EEV |
|---|---|---|
| 4000 | 3.2% | 3.9 % |
| 5500 | 3.4 % | 3.4 % |
| 7000 | 3.3 % | 5.7 % |
| 9000 | 3.7 % | 5.2 % |

5.5. Quantum efficency

The CCD quantum efficiency (QE) was measured in the wavelength interval 2000-10500 Å in incremental steps of 500 Å. The CCD is illuminated using monochromatic light obtained through a Xenon lamp, a series of filters and a monochromator. The signal integrated on the CCD is then compared to the response of a calibrated photodiode. The major contribution to the absolute error on the QE is given by the calibration error of the photodiode that ranges from 5% to 7% depending on the wavelength. Another source of error is the accuracy of the system gain measurement. Finally, one has to take into account the fact that the image of the monochromator slit on the CCD is usually small compared to the dimensions of the whole sensitive area making the non-homogeneity of the CCD another source of inaccuracy. Figure 3 shows the QE of the two CCDs LORAL and one of the two EEV4280. It is interesting to note that the QE of the two LORAL CCDs peaks at two different wavelengths, i.e. 6000 Å for the LRS CCD (Dolores) and 4000 Å for the OIG CCD. This is due to the different thickness of the anti-reflection ($Hf_2O$) coating which is 700 Å in the case of the LRS CCD and 500 Å for the OIG CCD.

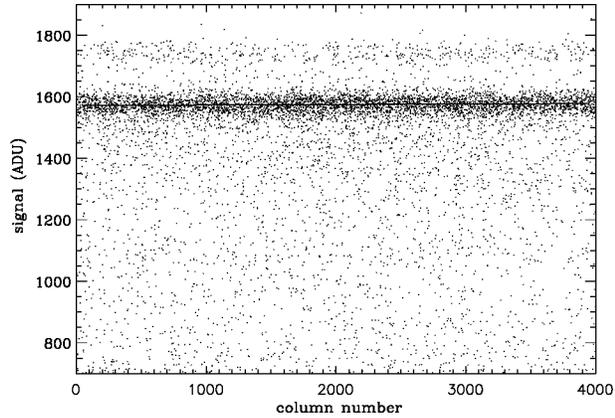

Fig. 4. histogram of the single pixel x-ray events for one of the EEV CCDs (OIG)

### 5.6. Dark current

The dark current is measured taking a short dark exposure and a long dark exposure. The signal level in the two exposures is then compared. The frame is divided in 10X10 pixel subframes and the dark current is computed as the average of the values measured in each of the subframes. In the case of the EEV 4280 the average value of the dark current turned out to be 6 e-/pix/hour.

### 5.7. Charge Transfer Efficiency (CTE)

The charge transfer efficiency (CTE) is measured using the X-ray stimulation method. In particular, the parallel CTE is obtained exposing the CCD to the $Fe^{55}$ source for a given amount of time. After integration, the columns are stacked together and the signal is plotted versus the number of pixel transfers. The CTE is then given by:

$$CTE = 1 - \frac{(Chargeloss)}{1620}\frac{e}{DN}$$

One example of the results obtained is shown in figure 4 for a LORAL (OIG) chip.

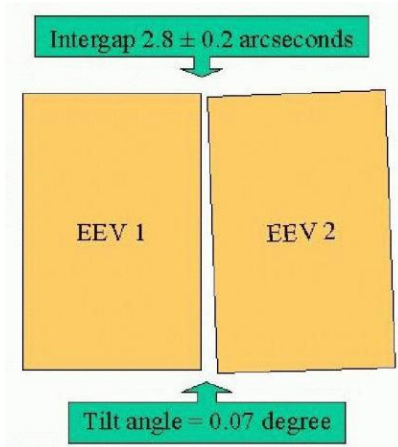

Fig. 5. The two CCD EEV 4280 mounted in OIG

The LORALs GTE turned out to be 0.999982 while the EEVs exhibited a GTE of 0.99999.

6. Test at the TNG

The tests at the telescope have been performed using the optical imager. We have chosen two EEV 4280, placing them one close to the other, to form a 4Kx4K mosaic. The GGD Mosaic have an intergap of 2.8 ± 0.2 arcseconds and a tilt angle of 0.07 degree (fig. 5 ).

Tab. 3 - Color equation

| EEV 1 | EEV 2 |
|---|---|
| U - u = 24.94 + 0.10 * ( U - B) | U - u = 25.05 + 0.11 * (U - B) |
| B - b = 26.63 + 0.10 * (B - V) | B - b = 26.71 + 0.11 * (B - V) |
| V - v = 26.38 - 0.06 * (B - V) | V - v = 26.42 - 0.05 * (B - V) |
| R - r = 26.28 - 0.11 * (V - R) | R - r = 26.29 - 0.15 * (V - R) |
| I - i = 25.45 + 0.07 * (R - I) | I - i = 25.45 + 0.10 * (R - I) |

The color equations, are showed in tab. 3.
Figure 6 show on the left a color picture of M57 ring nebula in Lyrae, made from B,V,R OIG frames, with an exposure time of 60 seconds and on the rigth a color composite picture of planetary nebula NGG 40, made with U,V and R OIG frames. The seeing is 0.85". The exposure times are 600 seconds in U and 180 seconds in V and R.
In figure 7 a color picture of SN 1999cl in NGG 4501, made from B,V,R OIG frames. The SN is located 46"W, 23"N with respect to the galaxy center.
Figure 8 show the center region of NGG 2903, made from B, V, R frames.

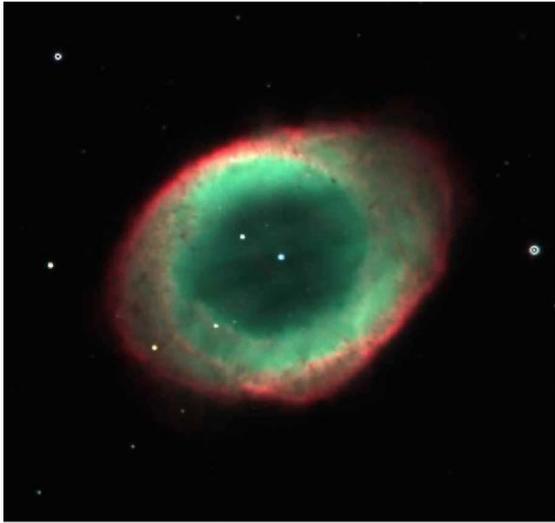 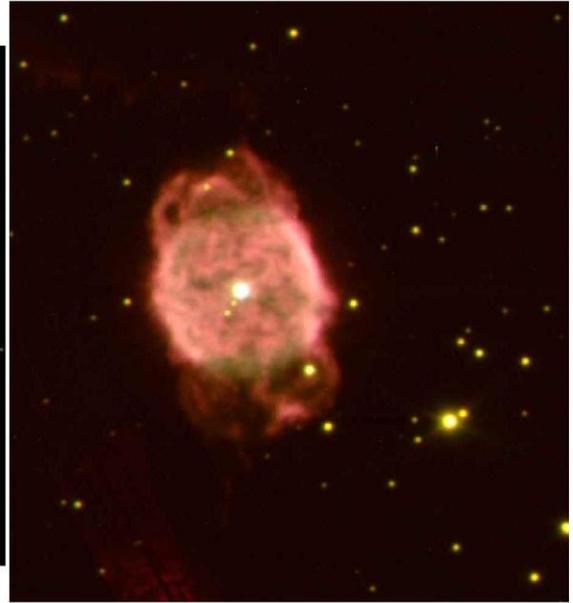

Fig. 6. On the left color picture of M57 ring nebula in Lyrae, on the right color composite picture of planetary nebula NGC 40.

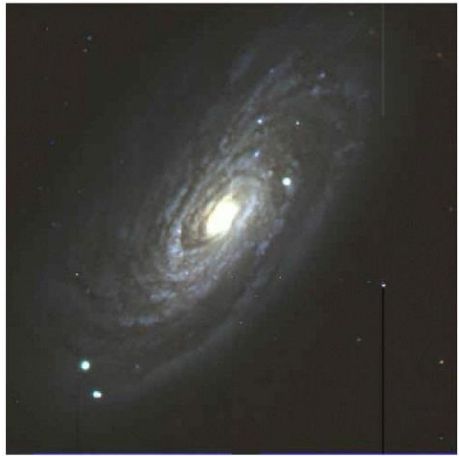

Fig. 7. Color picture of SN 1999cl in NGC 4501.

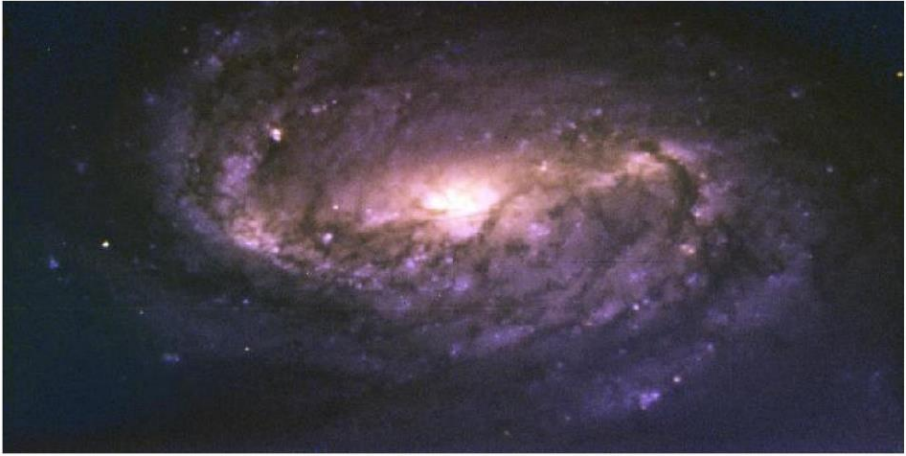

Fig. 8. The color composition of NGC 2903

Acknowledgements

Based on observations made with the TNG telescope operated on the island of La Palma by the Centro Galileo Galilei in the Spanish Observatorio del Roque de los Muchachos of the Instituto de Astrofisica de Canarias. We thanks the staff of TNG for the data reduction of the images taken at the telescope.